\documentclass[12pt]{article} 
\usepackage{epsfig}

\newcommand{\beq}{\begin{equation}}
\newcommand{\eeq}{\end{equation}}
\newcommand{\beqa}{\begin{eqnarray}}
\newcommand{\eeqa}{\end{eqnarray}}
\newcommand{\om}{\Omega_m}
\newcommand{\omh}{\Omega_m h^2}

\newcommand{\ap}{Alcock-Paczy{\`n}ski}

\def\ga{\mathrel{\mathpalette\fun >}}
\def\fun#1#2{\lower3.6pt\vbox{\baselineskip0pt\lineskip.9pt
  \ialign{$\mathsurround=0pt#1\hfil##\hfil$\crcr#2\crcr\sim\crcr}}}

\begin{document} 
\begin{center} 
{\bf Baryon Oscillations as a Cosmological Probe}
\vspace{0.1in} 

Eric V.~Linder\\ 
Physics Division, Berkeley Lab
\vspace{0.2in}

\end{center} 
\vspace{0.3in} 

Mapping the expansion of the universe gives clues to the underlying physics  
causing the recently discovered acceleration of the expansion, 
and enables discrimination among cosmological models.  We examine the 
utility of measuring the rate of expansion, $H(z)$, at various epochs, 
both alone and in combination with distance 
measurements.  Due to parameter degeneracies, it proves most useful as a 
complement to precision distance-redshift data.  Using the baryon 
oscillations in the matter power spectrum as a standard rod allows 
determination of $H(z)/(\omh)^{1/2}$ free of most 
major systematics, and thus provides a window on dark energy properties.  
We discuss the addition of this data from a next generation 
galaxy redshift survey such as KAOS to precision distance information from a 
next generation supernova survey such as SNAP.  This can provide useful 
crosschecks as well as lead to improvement on estimation of a time variation 
in the dark energy equation of state by factors ranging from 15-50\%. 

\vspace{0.1in}

\section{Introduction \label{sec.intro}} 

We now have strong evidence that the expansion of the universe is 
accelerating, 
from the original method of Type Ia supernova distance-redshift measurements 
\cite{perl99,riess98} and concordant observations of the cosmic microwave 
background (CMB) power spectrum and of large scale structure 
\cite{spergel,perc}.  The nature of the dark energy responsible 
for the acceleration will have profound 
implications for cosmology, particle physics, and fundamental physics.  
Mapping 
the expansion history of the universe offers a way to gain insights into the 
dark energy and the fate of the universe, for example by characterizing the 
equation of state behavior that is directly related to properties of 
the scalar field potential. 

As discussed in Linder \cite{lin0212}, one would like to carry out 
this mapping 
with not only precision measurements of the distance-redshift 
relation, but ideally 
with data on differential distances corresponding to the change 
between neighboring 
redshift epochs. The former, notably from the Type Ia supernova 
method, have proved 
adept at constraining the energy density and equation of state of 
the dark energy, 
with great improvements expected in the next decade.  But these 
involve an integration 
over the expansion rate behavior $H(z)$, which itself involves a 
redshift integral 
over the equation of state $w(z)$.  Probes more closely related to 
the differential 
distance might give $H(z)$ more directly.  

However the integral nature of the 
distance-redshift relation also provides the power to break 
degeneracies between 
cosmological parameters, which is an equally important aspect.  
So \cite{lin0212} 
found that the \ap\ effect of the cosmic shear distortion -- due to 
the source 
distances radial and transverse to the line of sight being measured 
at different 
epochs -- did not in fact automatically give more stringent estimations of 
the dark energy 
properties, despite involving a bare factor $H(z)$.  The cosmic shear 
(not to be confused with the local, 
weak lensing shear) is related to the ratio of the 
differential distance 
over some redshift interval to the integrated distance to the 
source.  So it is 
interesting to consider whether the situation changes if we can 
independently 
measure the two quantities, basically finding the Hubble parameter 
$H(z)$ separately. 

In Section \ref{sec.hz} we investigate the use of $H(z)$ for constraining 
the cosmological model.  But in Section \ref{sec.bo} we find that the 
most promising technique -- the baryon oscillation method -- actually 
measures a slightly different quantity.  We then examine the use of the 
radial and transverse distances provided by precision next generation 
galaxy redshift survey observations of the linear matter power spectrum, 
separately and together.  In Section \ref{sec.snbo} we show that the 
full power of the method comes from adding the information to a deep 
distance survey such as from accurate observations of Type Ia supernovae 
(e.g.~SNAP).  We summarize our conclusions and the need for future work 
in Section \ref{sec.concl}. 

\section{Using $H(z)$ Information\label{sec.hz}}

In this section we consider a data set giving the Hubble parameter $H(z)$ at 
some redshifts $z$, with a certain fractional precision.  This is a purely 
theoretical investigation as we do not specify how the measurements are made. 
Indeed, as mentioned in the introduction, the cosmic shear method only gives 
the product of $H(z)$ with the distance corresponding to the redshift $z$, 
and as we will see in \S\ref{sec.bo} the baryon oscillation method also 
provides a ratio involving $H(z)$.  So this is meant as a thought experiment. 

Similarly, it is obvious that knowledge of $H(z)$ over the entire redshift 
range from the observer at $z=0$ out to some depth is overly optimistic and 
would supersede any distance measurements in that range.  So we consider 
data at only a few redshifts in a narrow range and ask what cosmological 
information this can provide and what value it adds to a more realistic set 
of distance measurements.  Recall that the comoving distance $r$ or 
conformal time $\eta$ is related to $H(z)$ in a flat universe by 
\beq
r(z)=\eta(z)=\int_0^z dz'\,H^{-1}(z'), \label{rz}
\eeq 
and the angular diameter distance $d_a=(1+z)^{-1}r$ and luminosity distance 
$d_l=(1+z)\,r$.  The differential distance along the line of sight (radially) 
is simply $dr_\parallel=d\eta=dz/H$ and transversely is $dr_\perp=\eta\theta$, 
where $\theta$ is the angle subtended. 

Through the Friedmann equations, the expansion rate $H(z)$ is related to the 
cosmological components by 
\beq 
(H/H_0)^2=\om(1+z)^3+(1-\om)e^{3\int_0^z d\ln(1+z')\,[1+w(z')]}, \label{fried} 
\eeq 
where $H_0$ is the Hubble constant, the present value $H(z=0)$, $\om$ is the 
dimensionless matter density, and $w(z)$ is the equation of state of the dark 
energy.  We can examine the impact of measurements of $H(z)$ on determinations 
of the cosmological parameters through the sensitivities $\partial H/\partial 
\om$ etc., achieving formal constraints through the Fisher matrix method 
\cite{fisher}. 

\begin{figure}[!hbt]
\begin{center} 
\psfig{file=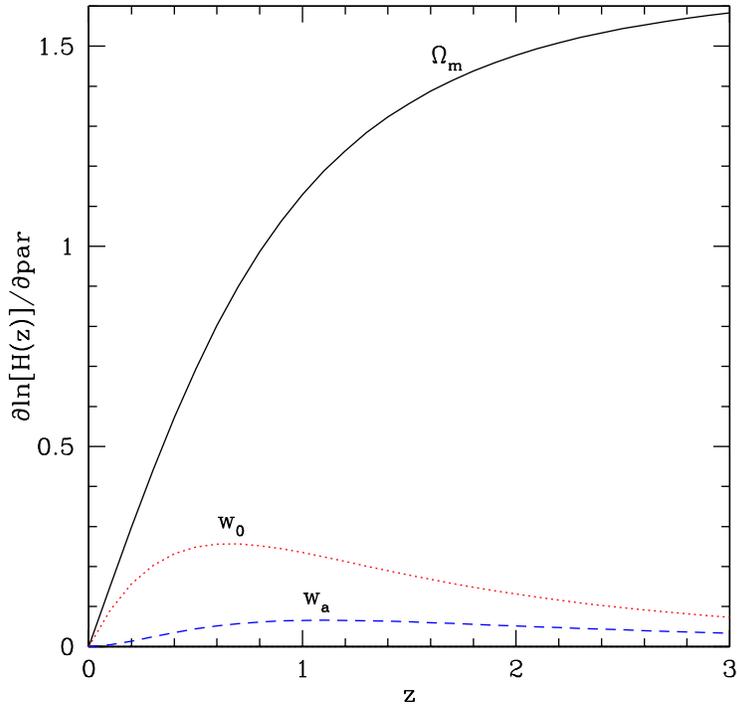,width=4in} 
\caption{The sensitivity of the expansion rate $H(z)$ to the 
cosmological parameters is plotted as a function of redshift. 
The larger the derivative at a particular redshift, the more 
constraining the observations are there, but the curves by 
themselves do not account for degeneracies between the parameters. 
} 
\label{fig.hz}
\end{center} 
\end{figure}

The sensitivities are shown in Figure \ref{fig.hz}, with the parametrization 
$w(z)=w_0+w_az/(1+z)$ of Linder \cite{lin0208} that allows robust treatment 
of the equation of state to redshifts greater than one.  However, 
the sensitivities are not the whole story: degeneracies 
between the parameters play a major role in determining them.  For example, 
while a 1\% measurement of $H(z)$ at $z=3$, say, would apparently constrain 
$\om$ to 0.06, $w_0$ to 0.14, and $w_a$ to 0.3, this holds only upon 
fixing all parameters but one.  In fact a measurement at a 
single redshift only gives an infinite ellipsoid in the joint 
three dimensional 
parameter space.  Even over a redshift range, such as $H(z)$ to 1\% at 
$z=2.8$, 3, 3.2, the uncertainties are uselessly large: $\sigma(\om)=0.87$, 
$\sigma(w_0)=76$, $\sigma(w_a)=207$.  But because the ellipsoid is 
fairly narrow, 
and the degeneracy direction is different than for distance measurements, 
the {\it} combination of $H(z)$ information with distance information can 
be valuable.  

For example, adding the estimation of $H(z)$ at $z=2.8$, 3, 3.2 
to a simulation of the data expected from the Supernova/Acceleration Probe 
(SNAP; \cite{snap}) survey out to $z=1.7$ allows parameter determination to 
$\sigma(\om)=0.0082$, $\sigma(w_0)=0.078$, $\sigma(w_a)=0.45$.  This 
represents a 
factor 3.5 improvement in constraining $\om$, 2\% in $w_0$, and 23\% in the 
measure of the time variation $w'\approx w_a/2$, relative to the canonical 
SNAP results.  So as expected there is clearly value in obtaining measurements 
of $H(z)$ (though we have not established how such would be carried out) -- 
though only in complementarity with a distance probe. 

Indeed one can show that measurements of $H(z)$ at redshifts $z>1$ basically 
act like information about the matter density $\om$.  If one eschewed any 
$H(z)$ data but added a prior $\sigma_{\Omega_m}=0.0082$ to the SNAP 
data then one 
would roughly recover the previous parameter estimations.  This is 
not surprising since at $z>1$ one is increasingly in the matter dominated, 
deceleration epoch and the expansion rate therefore best measures the matter 
density, not the dark energy properties.  So an integral measure such as the 
distance-redshift relation actually has an advantage in probing the dark 
energy equation of state, despite this quantity entering the distance 
through a double integral. 

We emphasize this important point further by considering two elaborations. 
If we spread the redshift range of the $H(z)$ measurements, to $z=2.5$, 3, 3.5 
{\it and} add simulated information from the future Planck cosmic microwave 
background survey \cite{planck}, then the dark energy constraints are 
still weak: $\sigma(\om)=0.039$, $\sigma(w_0)=1.6$, $\sigma(w_a)=5.4$.  
Again, the CMB has limited sensitivity to the dark energy equation of 
state and little complementarity with the $H(z)$ measurement.  
If we now add the SNAP data, the estimations improve to 0.0056, 0.070, 
and 0.34 respectively, but little of this is due to $H(z)$ since the 
CMB complementarity is much stronger.  The part of the improvement 
due to $H(z)$ is mostly restricted to $\om$ (since that is what $H(z=3)$ 
best probes) and somewhat $w_0$ (due to its degeneracy with $\om$); 
adding $H(z)$ tightens estimation of $\om$ by 44\%, $w_0$ by 11\%, 
but $w_a$ by only 4\%. 

\begin{figure}[!hbt]
\begin{center} 
\psfig{file=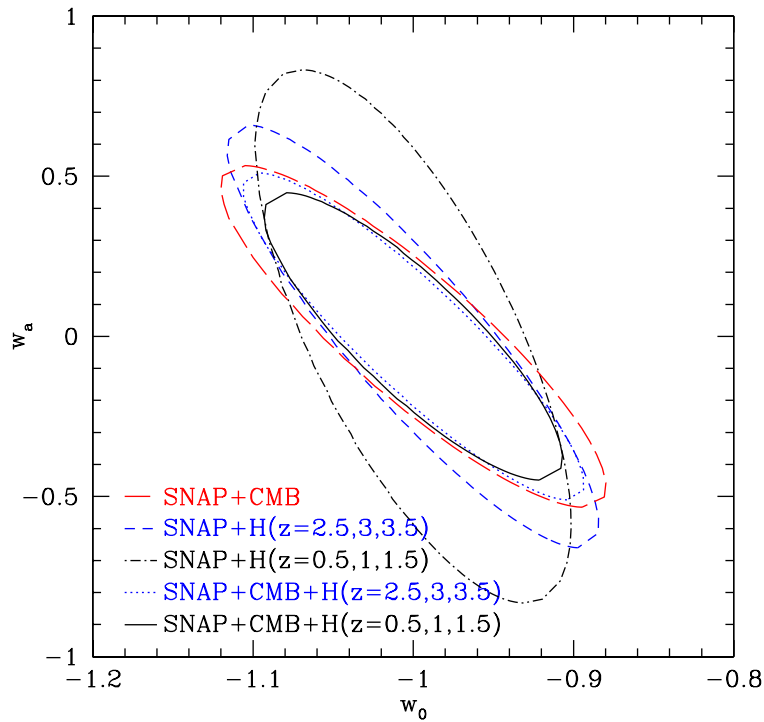,width=4in} 
\caption{Joint probability contours (68\% confidence level) for the 
time variation $w_a$ and the present value $w_0$ of the dark energy 
equation of state show that measurement of $H(z)$ is not as powerful 
as the CMB in complementing supernova data. 
} 
\label{fig.snboh}
\end{center} 
\end{figure}

For determination of $H(z)$ near $z=1$, the situation is only slightly better. 
It no longer acts as predominantly a matter density prior, but 
again $H(z)$ by itself cannot constrain the dark energy parameters, even 
with observations over a range $z=0.5-1.5$.  Furthermore, it has less 
complementarity with SNAP data and improves $w_a$ constraints by only 6\%.  
But conversely it gains in complementarity 
with the CMB information, and improves SNAP+CMB parameter determinations 
of $\om$, $w_0$, $w_a$ by 12\%, 23\%, 16\% respectively.  These 
various cases are illustrated in Fig.~\ref{fig.snboh}. 

So the distance data plays a central role in determining dark energy 
properties and $H(z)$ measurements only a 
subsidiary, complementary one.  But in any case we note that we have 
not identified any cosmological probe that provides these Hubble 
parameter measurements, let alone at 1\% accuracy. 

\section{Baryon Oscillations \label{sec.bo}} 

One cosmological probe that shows promise in obtaining differential distance 
measurements is the use of the imprint of the primordial baryon-photon 
acoustic oscillations in the matter power spectrum.  These are analogous 
to the oscillations appearing in the CMB temperature power spectrum, but 
are much smaller, appearing as wiggles superposed on the larger dark matter 
component (see \cite{husug,eishu} for a comprehensive discussion).  The 
wiggle wavelength can be used as a standard ruler, since the intrinsic 
scale is known from well understood physics of the matter-radiation 
decoupling epoch.  Then the angular or redshift space scale can be measured 
through a wide field redshift survey (though beyond the current state of 
the art) and the comparison probes the cosmological model.  

By observing at redshifts $z>1$ some of the wiggles appear in the linear 
density regime of the power spectrum, and by using only the locations 
and not the amplitudes of the oscillations one does not require 
problematic models of structure formation and evolution.  This method 
then has several positive aspects: simple, 
linear physics free from astrophysical uncertainties, direct relation of 
observations to cosmological quantities, and sensitivity to a snapshot of 
the expansion rate, $H(z)$.  For further discussion of the details and 
possible implementation of this probe see \cite{eis,seoeis,blakeglaze}. 

However, the baryon oscillations do not provide a pure measure of $H(z)$. 
Rather, the physics involves the ratio of the ``standard rod'' size  
to the observed oscillation scale (generally in Fourier wavenumber, 
$k$-space).  So the central quantity is 
\beq 
K \equiv {k_A\over k_{obs}}={1\over s}dz{d\eta\over dz}={dz\over H(z)\,s},  
\label{ka}\eeq 
where $k_A$ is the predicted acoustic oscillation scale, proportional to 
the inverse of the sound horizon $s$, and $k_{obs}$ is the observed scale, 
proportional to the inverse of the standard rod length $d\eta$. 
The sound horizon is given by 
\beqa 
s &=& \int_{z_{dec}}^\infty dz\,(c_s/H) \label{eq.s}\\ 
&=& (\omh)^{-1/2}\int da\,c_s\,\Bigr[a+a_{eq}+(1-\om^{-1})a^4e^{-3\int d\ln 
a\,[1+w(a)]}\Bigl]^{-1/2}, 
\eeqa 
where $a=(1+z)^{-1}$ is the scale factor of the universe, $c_s$ is the 
sound speed in the baryon fluid, $z_{dec}$ is the redshift of decoupling, 
and $a_{eq}$ is the scale factor at matter-radiation equality.  Note that 
while a cosmological constant ($w=-1$) would cause the last term in the 
brackets to have a negligible contribution 
to the integrand, some forms of dynamical dark matter could have  
non-negligible influence at these redshifts (see \cite{caldwell} 
for a discussion of early quintessence).  The sound speed for 
adiabatic perturbations in the baryons is 
\beq 
c_s={1\over\sqrt{3}}\left(1+{3\over 4}{\rho_b\over\rho_\gamma}\right)^{-1/2}. 
\eeq
Since the baryon density $\rho_b\sim\Omega_b h^2$ is well determined by 
current CMB measurements, and will be further improved by Planck data, 
and the photon density $\rho_\gamma\sim T_\gamma^4$ is also accurately 
known, then we can regard $c_s$ as fixed. 

From the form of Eqs.~(\ref{ka}) and (\ref{eq.s}), we see that $H(z)$ enters 
in both numerator and denominator, as itself and as an integrand.  This 
is the same form as for the cosmic shear probe \cite{lin0212}.  So as 
pointed out there, the value of the Hubble constant $H_0$ or $h$ 
does not enter the problem and therefore does not require marginalization. 
This is a definite advantage.  Furthermore, one can divide numerator 
and denominator by $(\omh)^{1/2}$.  At a casual glance, one might 
think that all 
dependence on this quantity is then removed.  But in fact, the approximation 
that $s\sim (\omh)^{-1/2}$ is not a good one, as pointed out in \cite{lin97}. 
There, a closer approximation for a flat, cosmological constant universe 
was found to be $s\sim (\omh)^{-0.3}$, while a more precise analysis 
\cite{eishu} is equivalent to $s\sim (\omh)^{-0.23}$.  The additional 
factors come from the presence of $a_{eq}$ and to a much lesser extent 
$a_{dec}$.  However, since the dependence arises from the sound horizon, 
it is the same for all the redshifts at which the measurements of 
$k_{obs}$ are carried out by the redshift survey.  This, combined with 
the precision to which Planck will determine $\omh$, means 
that its uncertainty couples very weakly to the other parameters (this was 
explicitly tested), and we will neglect it.  However, one still must 
incorporate the uncertainty in $\om$ in order to obtain realistic 
parameter error estimations. 

Therefore, the baryon oscillation method can essentially provide measurements 
of two cosmological variables according to Eq.~(\ref{ka}): $\tilde H(z)\equiv 
H(z)/(\omh)^{1/2}$ and $\tilde\eta(z)\equiv r(z)\,(\omh)^{1/2}$.  These come 
respectively from the wavenumbers along ($d\eta$) and transverse ($\eta$) 
to the line of sight.  This distinction from a plain $H(z)$ as treated in 
\S\ref{sec.hz} is important for the parameter degeneracies and complementarity 
with other methods. 

\begin{figure}[!hbt]
\begin{center} 
\psfig{file=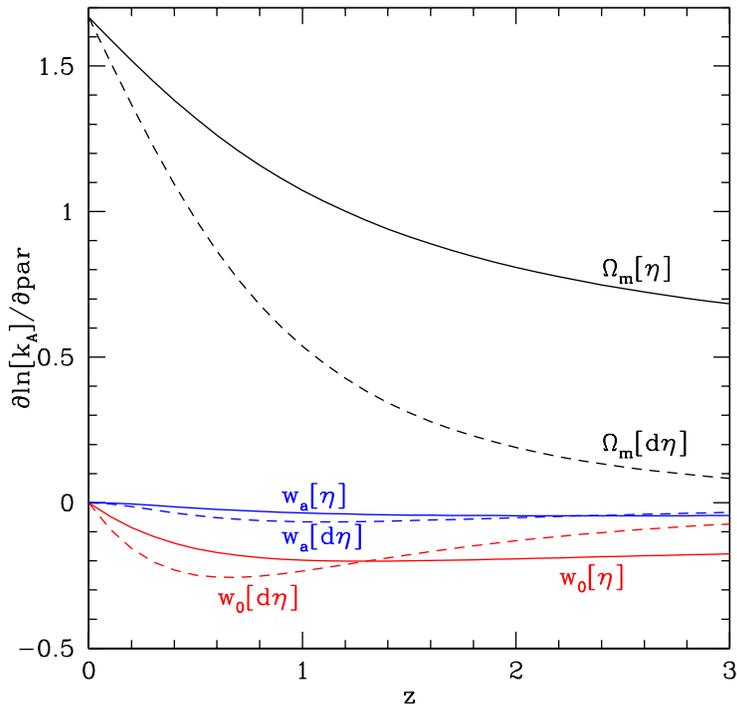,width=4in} 
\caption{As Fig.\ \ref{fig.hz} but for the baryon oscillation scale. 
Baryon oscillation measurements provide two quantities, corresponding 
to wave modes along [$d\eta$] and transverse [$\eta$] to the line of 
sight.  The $\om$ curves do not have zero sensitivity at $z=0$ 
because of the division by $(\omh)^{1/2}$. 
} 
\label{fig.bosens}
\end{center} 
\end{figure}

Figure \ref{fig.bosens} shows the sensitivities for these two quantities. 
As expected, for $w_0$ and $w_a$ the derivatives are the same as for $H(z)$ 
and $r(z)$.  However, the degeneracy relations between the parameters 
have now changed, and so the strength of the estimations have as well.  
Again we find that the probe in isolation cannot effectively constrain 
the cosmological model -- 
even the matter density since most of its dependence has been removed in 
the ratio.  Even in combination with CMB data it has little leverage. 

However the situation changes significantly for a fiducial model that 
has time variation in the equation of state.  For a supergravity inspired 
model \cite{braxm} that is well fit by $w_0=-0.82$, $w_a=0.58$, 
the oscillation 
data offers definite sensitivity to the time variation.  Now a 2\% (1\%) 
measurement of $K$ in both its radial and transverse aspects, in combination 
with Planck data, allows estimation of $w_a$ to 0.29 (0.20) for 
measurements at $z=\{0.5, 1, 1.5\}$ and 0.62 (0.47) for $z=\{2.5, 3, 3.5\}$.  
However, the 
estimations of $w_0$ remain poor: 0.16 (0.09) and 0.36 (0.27) respectively. 
Furthermore, since only 1-2 wiggles are in the linear regime at $z\approx1$, 
the observations are unlikely to achieve better than 2\% precision there, 
while we see 
that even 1\% precision at $z\approx3$ gives less impressive results.  And 
of course we have no guarantee that the true cosmological model will have 
a strong time variation in the dark energy equation of state.  

\section{Baryon Oscillations Plus Supernovae\label{sec.snbo}} 

As with the $H(z)$ analysis in \S\ref{sec.hz}, the baryon oscillation 
method  cannot stand alone as a robust cosmological probe.  In this 
section we consider it in complement with a SNAP supernova distance 
survey.  As expected, we find that it does not behave in the same 
manner as $H(z)$, as effectively a prior on the matter density.  
In fact, unlike $H(z)$, within a cosmological constant model of dark 
energy the complementarity with precision distance measurements is 
now substantial, providing good 
constraints.  Adding oscillation information acts even slightly more 
strongly than adding CMB information, relative to supernovae. 
When the baryon oscillation information is added to SNAP+CMB, further 
modest improvements are seen -- around 14\% in both $w_0$ 
and $w_a$ for 2\% measurement 
of the oscillation scale and 30\% for 1\% measurement.  This is fairly 
insensitive to the exact redshift distribution of the matter power spectrum 
measurements, i.e.~for redshifts near 1 or 3, or a spread 
$z=\{0.5, 1, 1.5\}$ vs.\ $\{0.8,1,1.2\}$.  

\begin{figure}[!hbt]
\begin{center} 
\psfig{file=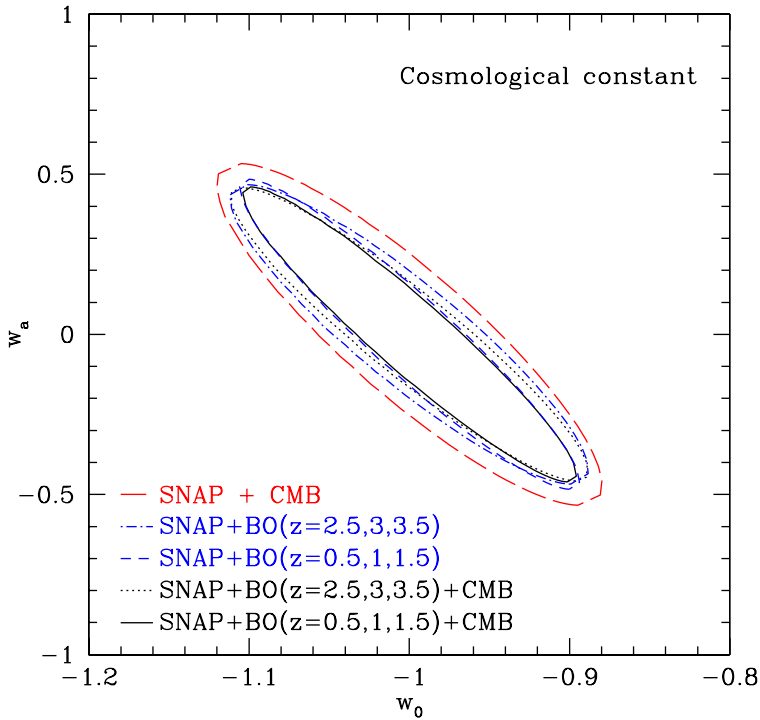,width=4in} 
\caption{As Fig.\ \ref{fig.snboh}, for baryon oscillation measurements 
and a cosmological constant model.  The baryon oscillation data is 
slightly stronger than the CMB data but not very complementary with it. 
} 
\label{snboc}
\end{center} 
\end{figure}

\begin{figure}[!hbt]
\begin{center} 
\psfig{file=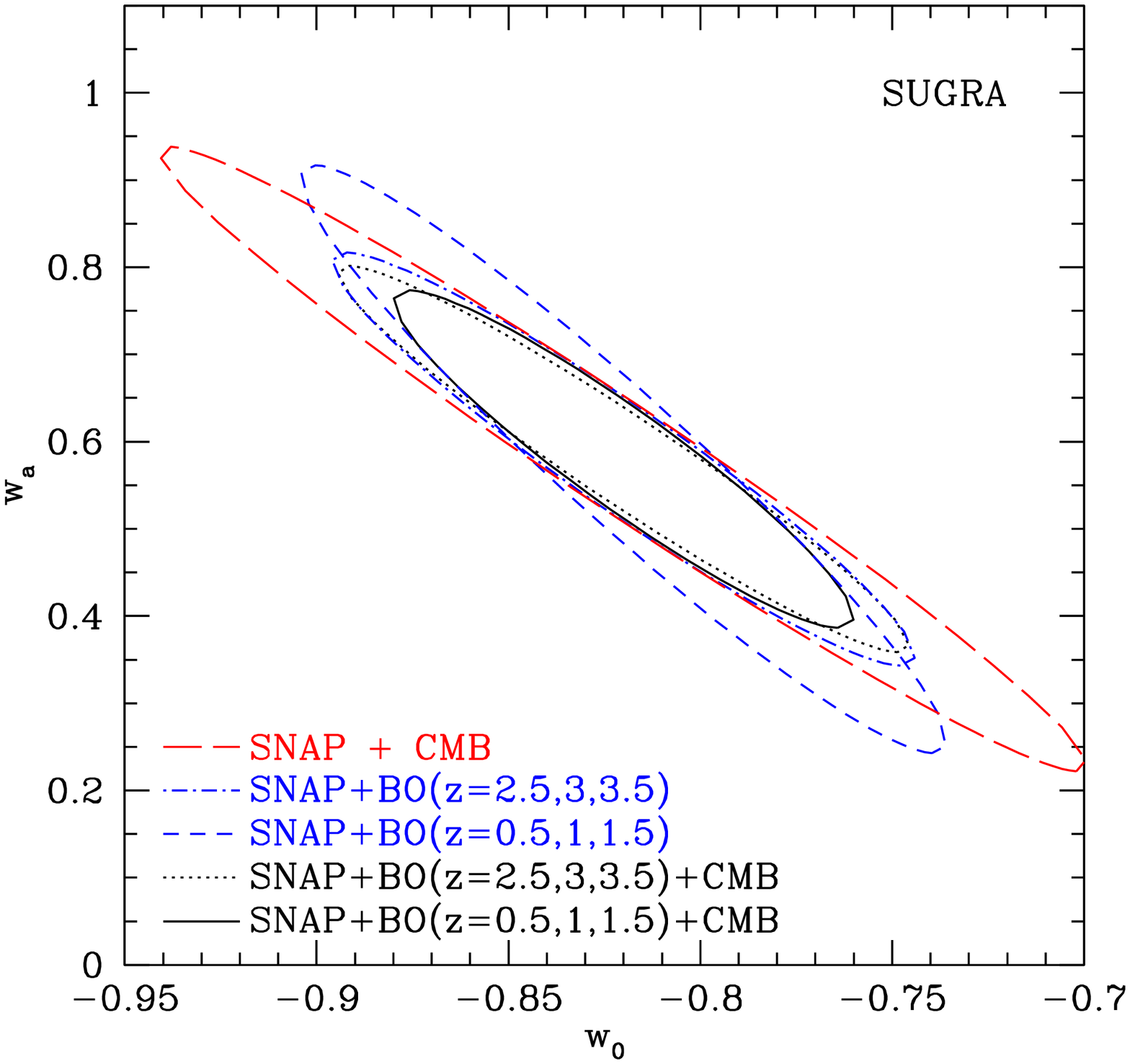,width=4in} 
\caption{As Fig.\ \ref{snboc} but for a supergravity inspired dark 
energy model.  Now baryon oscillation and CMB data are more complementary 
with each other as well as the supernova data, so the improvement with 
all three is more pronounced. 
} 
\label{snboc.sug}
\end{center} 
\end{figure}

For the SUGRA model, the improvement is stronger.   Baryon oscillations 
and CMB have increasing complementarity to each other and to supernovae. 
Now upon adding the oscillation probe to SNAP+CMB the estimation of 
$w_a$ sharpens by 46-37\% (54-60\%) for 2\% (1\%) precision, depending on 
whether the measurements 
are near $z=1$ or 3.  Furthermore, $\sigma(w_0)$ reduces by 51-39\% (59-57\%). 
So this offers hope that the baryon oscillation method can provide 
important complementarity useful in uncovering the nature of the dark 
energy.  The error contours for the cosmological constant and SUGRA cases 
are shown in Figs.~\ref{snboc} and \ref{snboc.sug}. 

Because of the insensitivity of the baryon oscillation results to the 
exact redshift range, so long as $z\ga1$, one can choose the survey 
characteristics based on observational considerations.  As mentioned, 
at redshifts $z<1.5$ one is likely to detect only 1 or 2 baryon wiggles, 
making it more difficult to precisely determine the oscillation scale 
$k_{obs}$.  But for $z>1.5$ the linear regime quickly extends in $k$-space, 
due to behavior of structure formation in a universe recently dominated 
by dark energy (see Fig.~1 of \cite{eis}), providing 3-4 detectable 
oscillations.  The redshift ranges most 
advantageous for observations are often identified as $z=0.5-1.3$ and 
$z=2.5-3.5$ \cite{blakeglaze} due to easy selection by 4000\AA\ break 
and Ly$\alpha$ features in the galaxies used in the survey.  Estimates of 
number of galaxies required and sky coverage are given in 
\cite{seoeis,blakeglaze}. 

Such a redshift survey could be accomplished by large telescopes on the 
ground within a decade.  One possibility is the KAOS project \cite{kaos}: 
the Kilo-Aperture Optical Spectrograph proposed as a front end for the 
Gemini South 8 meter telescope. This would have multiplexing capability 
from some 4000 fibers for simultaneous measurement of galaxy redshifts. 
With a 1.5 square degree field of view and coverage of some 400 square 
degrees of sky KAOS could measure precise redshifts for $10^6$ galaxies.  
This could provide estimates of the wiggle scale at the 2\% precision 
level \cite{seoeis}. 

Another intriguing idea is to use wide field observations from space. 
This would have the advantage of not being restricted to the $z\approx1$ 
and 3 
ranges just mentioned, which were limited by the Earth's atmosphere.  Indeed, 
from a theoretical point of view, a redshift range $z=1.5-2$ is essentially 
as powerful as $z\approx3$ in terms of number of oscillations mapped, a 
definite advantage over $z\approx1$, and yet requires less spectroscopic 
exposure time than the deeper survey.  Calculations show that the parameter 
estimation for a given precision is as tight as the lower or higher 
redshift ranges. 

While there is no planned 
massively multiplexing spectrograph for space, one interesting possibility 
is populating spare regions of the SNAP focal plane with grisms capable of 
low resolution spectroscopy.  Also, photometric redshifts can be 
generated with SNAP's nine filters.  There is no problem achieving 
the number or area statistics as the proposed SNAP wide field survey 
(mostly focused on weak gravitational lensing) will find $10^8$ galaxies 
over 300 square degrees. However with a 2 meter telescope and spectral 
resolutions of order 100 or less, this 
clearly is not capable of carrying out all the science that an 8 meter ground 
based telescope with high resolution spectrograph could.  Still, while 
this would not provide the same precision mapping of the 3D matter 
power spectrum, it might give decent quality information on the 2D, 
projected spectrum, roughly corresponding to the transverse wavenumber 
modes in Eq.~(\ref{ka}).  Photometric or low resolution spectroscopic  
redshifts would additionally give a smeared representation of the radial 
dimension.  Detailed analysis of the baryon oscillation method 
with SNAP is left for future work; here we simply investigate the parameter 
constraints from the transverse and radial modes separately. 

\begin{figure}[!hbt]
\begin{center} 
\psfig{file=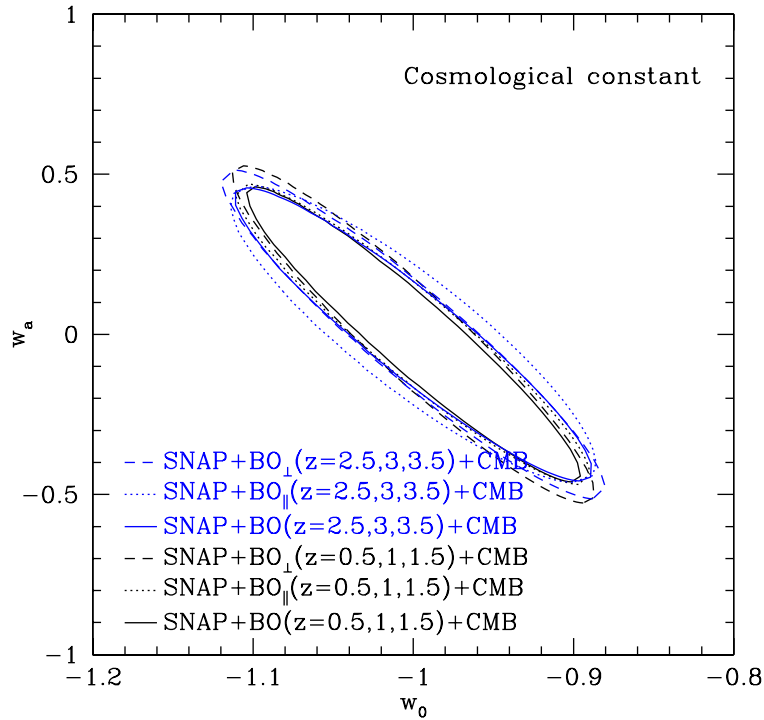,width=4in} 
\caption{As Fig.\ \ref{snboc} but separating the contributions from the 
baryon oscillation wave modes transverse and parallel to the line of 
sight.  Radial information, requiring accurate redshifts from a 
spectroscopic survey such as KAOS, is more useful. 
} 
\label{snboc.deta}
\end{center} 
\end{figure}

One expects that the radial mode, $d\tilde\eta$, which involves a bare factor 
$H(z)$, should provide better limits, while the transverse mode, $\tilde\eta$, 
acts basically like a distance-redshift measurement though with the important 
degeneracy differences previously mentioned.  Indeed the sensitivities plotted 
in Fig.~\ref{fig.bosens} bear this out (though the degeneracy 
relations are not there apparent; also note that the radial mode has 
low sensitivity at redshifts that are well into the matter dominated epoch).  
For example, denoting 
the full baryon oscillation information as BO, the radial only as 
BO$_\parallel$, and the transverse only as BO$_\perp$, one finds that 
2\% precision gives $\sigma(\om)=0.0057$, $\sigma(w_0)=0.069$, 
$\sigma(w_a)=0.30$ 
for SN+CMB+BO, (0.0073, 0.073, 0.31) for SN+CMB+BO$_\parallel$, and 
(0.0065, 0.075, 0.35) for SN+CMB+BO$_\perp$.  (Though presumably the precision 
in a full 3D survey would be better than in a 2D plus low resolution radial 
survey.) In the last case there is essentially no improvement over the SN+CMB 
case without any oscillation information. 
Various cases are illustrated in Fig.~\ref{snboc.deta}. 

\begin{figure}[!hbt]
\begin{center} 
\psfig{file=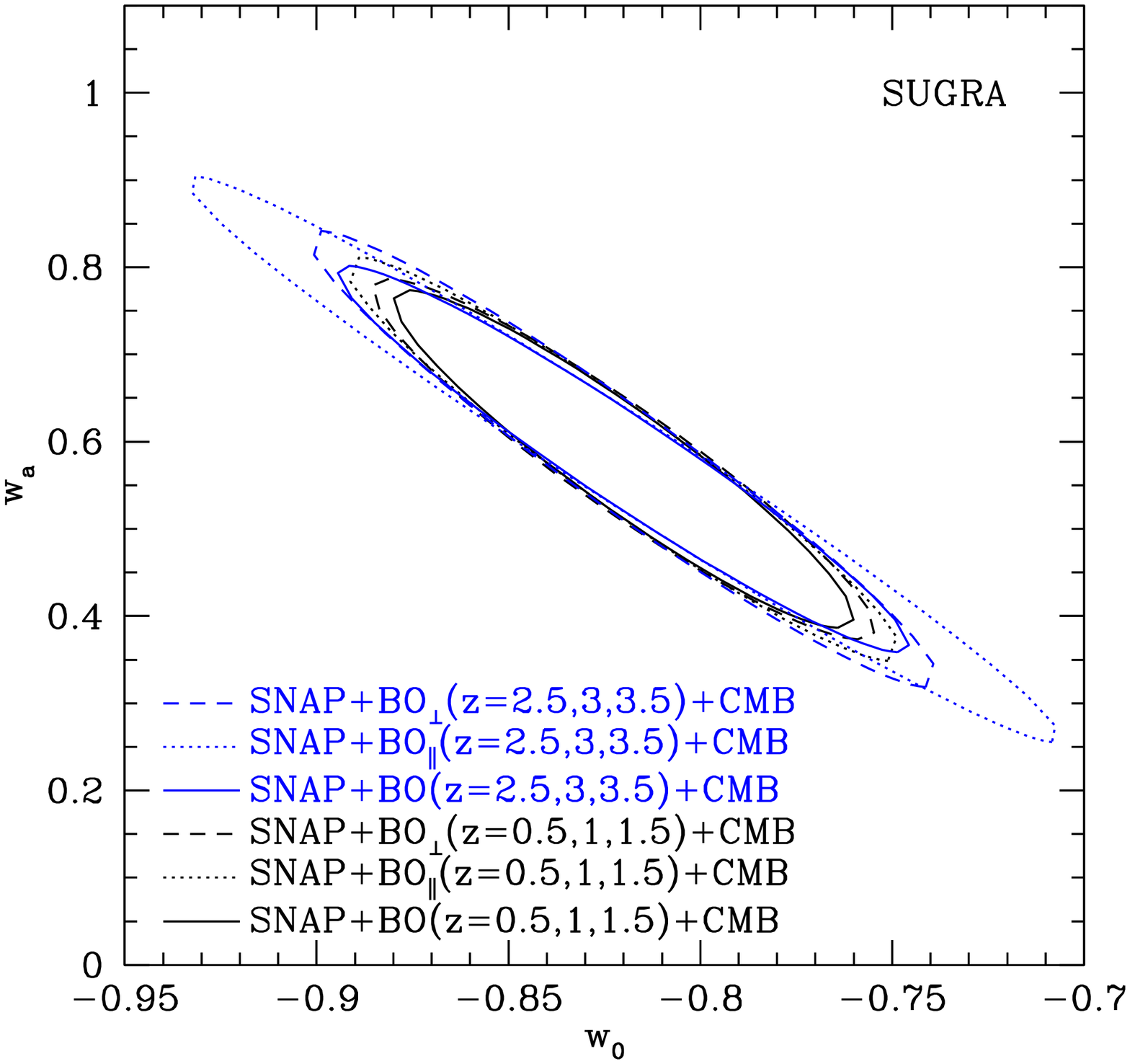,width=4in} 
\caption{As Fig.\ \ref{snboc.deta} but for a supergravity inspired 
dark energy model.  Now the transverse modes, detectable through a 
2D survey, are more influential. 
} 
\label{snboc.sug.deta}
\end{center} 
\end{figure}

Again, the baryon oscillation method is more useful in the presence of a 
time varying equation of state dark energy.  Moreover, due to the long 
baseline 
entering the integrated distance, the BO$_\perp$ information is actually more 
valuable than the radial, if at the same precision. The results for the SUGRA 
model are illustrated in Fig.~\ref{snboc.sug.deta}.  We see that 2\% precision 
data on the transverse modes adds complementarity to the supernova and CMB 
information, improving the estimates of $\om$ by 56\%, $w_0$ by 46\%, and 
$w_a$ by 42\%.  This represents the vast majority of the impact of the 
full baryon oscillation data discussed at the beginning of this section.

\section{Conclusion \label{sec.concl}}

Differential distance measurements, providing a snapshot of the expansion 
rate $H(z)$, have long seemed attractive theoretically as ways to probe 
the nature of dark energy.  But they have also appeared difficult to 
implement observationally.  The cosmic shear, or \ap, probe (not to 
be confused with promising weak lensing shear measurements) involves a 
ratio of differential to integrated distances, or the product $H(z)\,r(z)$, 
and \cite{lin0212} showed 
that it could act only in a minor, complementary role to precision 
distance observations.  The growth of structure in the linear regime 
also might be thought sensitive to $H(z)$ but at redshifts $z\ga2$ this 
essentially probes $\om$ not the dark energy; however, through 
other factors the linear growth 
still retains some sensitivity to the equation of state and 
its time variation \cite{jenklin}.  Nonlinear structure formation can 
involve $H(z)$ as a separate factor through the differential volume element 
in cluster or galaxy halo counts, but this is entangled in 
systematic uncertainties from nonlinear physics and observational selection 
effects \cite{mohr,levine}.  On large scales this may be ameliorated, 
but the needed numerical 
simulations of large scale structure incorporating a time varying equation 
of state are just now being carried out \cite{jenklin}. 

In this paper we have pushed these observational difficulties into the 
background and considered the use of $H(z)$ regardless.  Our conclusion is 
that it is not a panacea and only offers aid through complementarity with 
a deep, precision distance survey such as SNAP; then it contributes mild 
improvement to the cosmological parameter constraints.  This is basically 
due to $H(z)$ acting at higher redshifts as a determinant of the matter 
density, not a direct probe of dark energy properties.  At redshifts $z<1$ 
it has somewhat more leverage, but requires precision on the 1\% level 
for significant improvement. 

The baryon oscillation method of using wiggles in the matter power spectrum 
as a standard ruler determines a slightly different measure of the 
expansion rate $H(z)$.  This probe is sufficiently promising, though 
again only in complementarity with 
a supernova distance survey, that it should be pursued further.  
Depending on the nature of the dark energy, incorporation of oscillation 
measurements can offer significant improvements on estimation of the 
time variation of the equation of state.  One of the 
most striking aspects is its cleanness, based on simple, well understood 
physics and with no apparent major systematic uncertainties.  Note that in 
all the analyses presented here of different cosmological probes, only the 
SNAP data has included systematic uncertainties -- the $H(z)$ and baryon 
oscillation precisions have been taken as purely statistical.  For all 
known and other proposed probes this is certainly overly optimistic.  Whether 
systematics enter at the 1-2\% level in the baryon oscillation method, 
from, say, residual nonlinearities or mass vs.\ light bias, needs 
further investigation. 

Two interesting concepts for baryon oscillation observations are 
the KAOS project on the ground, and a spectroscopically less precise but 
reasonably straightforward space implementation with grisms or 
photometric redshifts from SNAP. 
We have seen that the optimal redshift range is not 
strongly determined by the parameter sensitivity, and so will be driven 
by trade offs in observation strategy.  Both projects deserve further 
investigation, though it is intriguing to imagine that SNAP could represent 
a cosmology superprobe -- incorporating the supernova distance, weak lensing, 
some part of the baryon oscillation, and possibly even cluster count 
methods of cosmological parameter determination.  But even if SNAP is 
rather promising for revealing the nature of dark energy, our 
understanding and confidence will still be strengthened by multiple, 
complementary and crosschecking next generation surveys.

\section*{Acknowledgments} 

I thank Greg Aldering, Matthew Colless, Daniel Eisenstein, Saul Perlmutter, 
and Martin White for useful discussions. This work has been supported 
in part by the Director, Office of Science, Department of Energy under 
grant DE-AC03-76SF00098.


\begin{thebibliography}{99}

\bibitem{perl99}
	S.\ Perlmutter {et al.},  Astrophys. J. {\bf 517}, 565 (1999)

\bibitem{riess98}
	A.\ Riess {et al.}, Astron. J. {\bf 116}, 1009 (1998)

\bibitem{spergel}
	D.N.\ Spergel et al., astro-ph/0302209

\bibitem{perc}
	W.J.\ Percival et al., MNRAS 337, 1068 (2002); astro-ph/0206256

\bibitem{lin0212}
	E.V.~Linder, astro-ph/0212301. 

\bibitem{fisher}
	M.\ Tegmark, D.J.\ Eisenstein, W.\ Hu, R.\ Kron, astro-ph/9805117

\bibitem{lin0208}
	E.V.~Linder, Phys.~Rev.~Lett.~90, 091301 (2003); astro-ph/0208512

\bibitem{snap}
	http://snap.lbl.gov ; G.~Aldering et al., in SPIE Proceedings 4835; 
	astro-ph/0209550 

\bibitem{planck}
	http://astro.estec.esa.nl/Planck

\bibitem{husug}
	W.\ Hu and N.\ Sugiyama, Ap.\ J.\ 471 (1996) 542

\bibitem{eishu}
	D.J.~Eisenstein and W.~Hu, Ap.~J.~496 (1998) 605; astro-ph/9709112

\bibitem{eis}
	D.~Eisenstein, in Proceedings from Wide-Field Multi-Object 
	Spectroscopy; astro-ph/0301623

\bibitem{seoeis}
	H.~Seo and D.~Eisenstein, in preparation 

\bibitem{blakeglaze}
	C.\ Blake and K.\ Glazebrook, astro-ph/0301632

\bibitem{caldwell}
	R.R.\ Caldwell et al., astro-ph/0302505 

\bibitem{lin97}
	E.V.~Linder, astro-ph/9712159

\bibitem{braxm}
	P.\ Brax and J.\ Martin, Phys.\ Lett.\ B468, (1999) 40 

\bibitem{kaos}
	http://www.noao.edu/kaos 

\bibitem{mohr}
	J.\ Mohr, http://www.xray.mpe.mpg.de/\~{}ringberg03/TALKS/Mohr.pdf

\bibitem{levine}
	E.S.\ Levine, A.E.\ Schulz, and M.\ White, Ap.\ J.\ 577 (2002) 
	569; astro-ph/0204273

\bibitem{jenklin}
	A.~Jenkins and E.V.~Linder, in preparation 

\end{thebibliography}
\end{document}